astro-ph/9605006  2 May 1996

# Are topological defects responsible for the 300 EeV cosmic rays?[*]

R.J. Protheroe[a] and P.A. Johnson[†a]

[a]Department of Physics & Mathematical Physics, University of Adelaide, SA 5005, Australia.

We use of a hybrid matrix–Monte Carlo method to simulate the cascade through the cosmic background radiation initiated by UHE particles and radiation emitted by topological defects. We follow the cascade over cosmological distances and calculate the intensities of hadrons, $\gamma$-rays and neutrinos produced. We compare our results with the observed cosmic ray intensity at 300 EeV and lower energies, and conclude that topological defects are most unlikely to be the origin of the most energetic cosmic ray events.

## 1. INTRODUCTION

Several calculations [1–6] of the propagation of UHE cosmic rays have been performed in which the sensitivity of the spectrum to various source scenarios is investigated. A general consensus has emerged that no particles should be detected above the Greisen-Zatsepin-Kuzmin (GZK) cut-off if they are produced at moderate extragalactic distances. However, at least three cosmic rays have been detected with energies above the GZK cut-off [7–9] We discuss the possibility that topological defects could be responsible for the "super-GZK events", i.e. cosmic rays observed above the expected GZK cut-off.

## 2. THE ORIGIN OF THE HIGHEST ENERGY COSMIC RAYS

The super-GZK events may result from a non-acceleration process. It has been suggested that they could be produced by the decay of super-massive X particles [10–13], themselves radiated during collapse or annihilation of topological defects, remnants of an early stage in the evolution of the Universe. The X particles have GUT-scale masses of order $10^{15} - 10^{16}$ GeV, and decay into leptons and quarks at lower energies. The hadronic content of particles from $X$ particle decay is thought to be $\sim 3\%$ nucleons and the rest pions with an $E^{-1.5}$ spectrum extending from $\sim 1$ GeV up to $\sim m_X c^2$ [13]. We inject this spectrum at various distances up to $z = 9$ and carry out a Monte-Carlo matrix propagation calculation as described in [6].

The mean interaction lengths and energy-loss distances for electrons, and mean interaction lengths for photons are given in Figs. 1 and 2 respectively. The intergalactic magnetic field we use is obtained by assuming magnetic flux conservation per co-moving area, i.e. $B(z) = B_0(1+z)^2$ where $B_0$ is the magnetic field adopted for the present epoch. In this paper we will generally use $B_0 = 10^{-9}$ gauss which is the at the upper end of estimates of the average extragalactic field [14], but we will discuss lower fields elsewhere.

The flux of "observable particles" (photons, electrons, protons and neutrons) at Earth multiplied by $E^2$ is plotted against energy and injection redshift in Fig. 3 assuming $B_0 = 10^{-9}$ gauss. The "ridge" in the spectrum at $10^9$ GeV is due to synchrotron radiation in the $B_0 = 10^{-9}$ gauss field by the first generation of electron-positron pairs (produced by double pair production on the microwave background or pair production on the radio background) which have energy $\sim 2 - 5 \times 10^{14}$ GeV. The flux above $10^{10}$ GeV is due to inverse-Compton emission, and the "valley" at $10^6$ GeV is due to photon-photon pair production on the microwave background. In Figure 4 we have integrated over redshift assuming uniform injection of the spectrum of nucleons and pions from topological defects (as described above) per co-moving volume, and normalized to the data at at 300 EeV. The $\gamma$-ray emission at 300 EeV and above is such that one would expect already to have detected approximately 7 events when only one has been observed. This appears to present serious problems for models where topological defects are responsible for the super-GZK events, because much of the emission from defects ends up in the electromagnetic channel with a hard spectrum extending up to GUT scale energies.

[*]The research of RJP is supported by a grant from the Australian Research Council.

[†]present address: DSTO, Salisbury, SA 5108, Australia

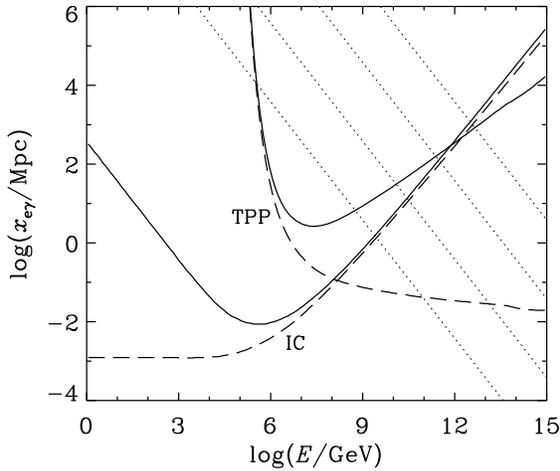

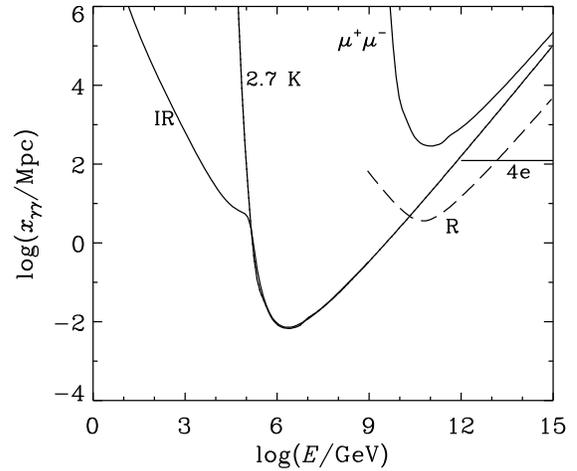

Figure 1. The mean interaction length (dashed line) and energy-loss distance (solid line), $E/(dE/dx)$, for electron-photon triplet pair production (TPP) and inverse-Compton scattering (IC) in the microwave background. The energy-loss distance for synchrotron radiation is also shown (dotted lines) for intergalactic magnetic fields of $10^{-9}$ (bottom), $10^{-10}$, $10^{-11}$, and $10^{-12}$ gauss (top).

Figure 2. The mean interaction length for pair production for $\gamma$-rays in the microwave background (2.7K), the infrared and optical background (IR), and the Radio Background (R). Also shown are the mean interaction length for muon pair production ($\mu^+\mu^-$) and double pair production (4e) in the microwave background.

## 3. CONCLUSIONS

We have made a new calculation of the propagation of the expected decay spectrum of topological defects through the extragalactic radiation fields over cosmological distances, and included in our work the calculation of secondary fluxes of $\gamma$-rays and neutrinos which result from interactions of the cosmic rays in the radiation field and subsequent cascading. Inclusion of synchrotron radiation in the cascade quantitatively confirms the prediction [12] that the expected $\gamma$-ray spec-trum will be dramatically higher. Our results present very serious problems for topological defect models which naturally produce a very hard injection spectrum of $\gamma$-rays with typically half the power going into $\gamma$-rays above $\sim 10^{14}$ GeV. This will be fully discussed in a future paper [16]

## REFERENCES

1. Hill C.T. and Schramm D.N. *Phys. Rev. D* **31** (1985) 564
2. Berezinsky V.S. and Grigor'eva S.I. *Astron. Astrophys.* **199** (1988) 1
3. Aharonian F.A. and Cronin J.W. *Phys. Rev. D* **50** (1994) 1892
4. Yoshida S. and Teshima M. *Prog. Theor. Phys.* **89** (1993) 833
5. Geddes J., Quinn T.C. and Wald R.M., preprint (1995)
6. Protheroe R.J. and Johnson P.A *Astroparticle Phys.* (1995) in press
7. Bird D.J. *et al. Ap. J.* **441** (1995) 144

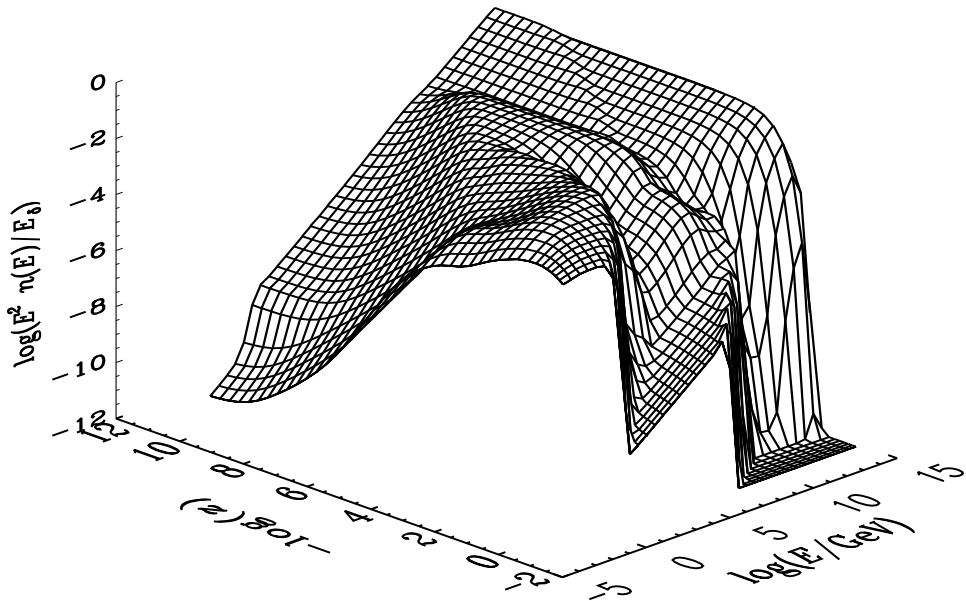

Figure 3. The flux of "observable particles" (photons, electrons, protons and neutrons) at Earth multiplied by $E^2$ is plotted against energy and injection redshift.

8.  Hayashida N. *et al.* *Phys. Rev. Lett.* **73** (1994) 3491

9.  Yoshida S. *et al.* *Astroparticle Phys.* **3** (1995) 105

10. Bhattacharjee P., Hill C.T. and Schramm D.N. *Phys. Rev. Lett.* **69** (1992) 567

11. Sigl G., Schramm D.N. and Bhattacharjee P. *Astroparticle Phys.* **2** (1994) 401

12. Bhattacharjee P. and Sigl G. *Phys. Rev. D* **51** (1995) 4079

13. Sigl G. to appear in *Space Sci. Rev.* (1995)

14. Kronberg P.P. *Rep. Prog. Phys.* **325** (1994) 382

15. Stanev T., in *Particle Acceleration in Cosmic Plasmas*, edited by G.P. Zank and T.K. Gaisser (American Institute of Physics, New York, 1992) p.379.

16. Protheroe R.J. and Stanev T.S. (1996) in preparation

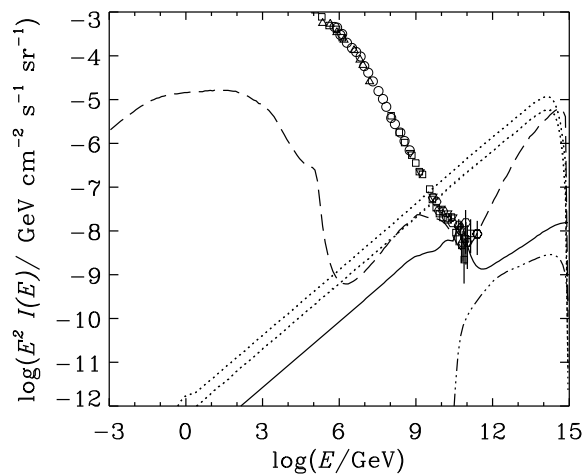

Figure 4. Intensity at Earth assuming uniform injection per co-moving volume of protons (solid line), neutrons (dot-dash line), photons (dashed line) and muon neutrinos (or antineutrinos) (upper dotted curve), and electron neutrinos (or antineutrinos) (lower dotted curve). Proton plus photon flux has been normalized to 300 EeV data [7]; other data shown is from [15].